\documentclass[aip,reprint]{revtex4-1}
\usepackage[utf8]{inputenc}

\usepackage{amsmath, amsthm, amssymb}
\usepackage{epsfig}
\usepackage{epstopdf}
\usepackage{hyperref}

\usepackage{ulem}

\usepackage{color}

\begin{document}

\title{Ultrafast Photodissociation Dynamics and Nonadiabatic Coupling Between Excited Electronic States of Methanol Probed by Time-Resolved Photoelectron Spectroscopy}

\author{Elio~G.~Champenois}
  \affiliation{Chemical Sciences Division, Lawrence Berkeley National Laboratory, Berkeley, CA 94720, USA}
  \affiliation{Graduate Group in Applied Science and Technology, University of California, Berkeley, CA 94720, USA}
\author{Loren~Greenman}
  \affiliation{Chemical Sciences Division, Lawrence Berkeley National Laboratory, Berkeley, CA 94720, USA}
  \affiliation{Department of Chemistry, University of California, Davis, CA 95616, USA}
  \affiliation{Department of Physics, Kansas State University, Manhattan, KS 66506, USA}
\author{Niranjan~Shivaram}
  \affiliation{Chemical Sciences Division, Lawrence Berkeley National Laboratory, Berkeley, CA 94720, USA}
\author{James~P.~Cryan}
  \affiliation{Stanford PULSE Institute, SLAC National Accelerator Laboratory, Menlo Park, CA 94025, USA}
\author{Kirk~A.~Larsen}
  \affiliation{Chemical Sciences Division, Lawrence Berkeley National Laboratory, Berkeley, CA 94720, USA}
  \affiliation{Graduate Group in Applied Science and Technology,
  University of California, Berkeley, CA 94720, USA}
\author{Thomas~N.~Rescigno}
  \affiliation{Chemical Sciences Division, Lawrence Berkeley National Laboratory, Berkeley, CA 94720, USA}
\author{C.~William~McCurdy}
  \affiliation{Chemical Sciences Division, Lawrence Berkeley National Laboratory, Berkeley, CA 94720, USA}
  \affiliation{Department of Chemistry, University of California, Davis, CA 95616, USA}
\author{Ali~Belkacem}
  \affiliation{Chemical Sciences Division, Lawrence Berkeley National Laboratory, Berkeley, CA 94720, USA}
\author{Daniel~S.~Slaughter}
\email[]{DSSlaughter@lbl.gov}
  \affiliation{Chemical Sciences Division, Lawrence Berkeley National Laboratory, Berkeley, CA 94720, USA}
\date{\today}

\begin{abstract}
The electronic and nuclear dynamics in methanol, following 156~nm photoexcitation, are investigated by combining a detailed analysis of time-resolved photoelectron spectroscopy experiments with electronic structure calculations.
The photoexcitation pump pulse is followed by a delayed 260~nm photoionization probe pulse, to produce photoelectrons that are analyzed by velocity map imaging. The yield of mass-resolved ions, measured with similar experimental conditions, are found to exhibit the same time-dependence as specific photoelectron spectral features. Energy-resolved signal onset and decay times are extracted from the measured photoelectron spectra to achieve high temporal resolution, beyond the 20~fs pump and probe pulse durations.
When combined with {\it ab initio} calculations of selected cuts through the excited state potential energy surfaces, this information allows the dynamics of the transient excited molecule, which exhibits multiple nuclear and electronic degrees of freedom, to be tracked on its intrinsic few-femtosecond timescale.
Within 15\,fs of photoexcitation, we observe nuclear motion on the initially bound photoexcited 2$^{1}$A$''$ (S$_2$) electronic state, through a conical intersection with the 1$^{1}$A$'$ (S$_3$) state, which reveals paths to photodissociation following C--O stretch and C--O--H angle opening.

\end{abstract}


\maketitle
\section{Introduction}

Molecules excited by light can evolve by coupled nuclear and electronic motion over ultrafast timescales.\cite{zewail_laser_1988} 
In many photochemical reactions, the products and their relative yields depend sensitively on the details of these coupled molecular dynamics and processes that occur within a few femtoseconds.\cite{peters_ultrafast_2017,timmers_coherent_2014,galbraith_few-femtosecond_2017,ludwig_ultrafast_2016,joalland_photochemical_2014,adachi_probing_2018}
For example, near a conical intersection, where the evolving molecular geometry finds two or more electronic states of equal potential energy, nonadiabatic transitions can occur, enabling important non-radiative relaxation mechanisms, such as competing isomerization or dissociation pathways.\cite{yarkony_diabolical_1996,levine_isomerization_2007}
Considerable interest in investigating such dynamics on their natural femtosecond timescales has driven recent developments of time-resolved spectroscopic methods employing few-cycle infrared lasers and broadband attosecond pulses.\cite{ergler_spatiotemporal_2006,nabekawa_sub-10-fs_2016,sandor_strong_2016,galbraith_xuv-induced_2017}

Time-resolved photoelectron spectroscopy (TRPES) has arguably become the gold standard to study ultrafast nonadiabatic dynamics in neutral excited molecules.\cite{stolow_femtosecond_2004,stolow_femtosecond_2003,neumark_time-resolved_2001}
A measurement of the instantaneous photoelectron binding energy of the excited wavepacket can be parsed into contributions from each state involved in the electronic relaxation process,\cite{satzger_reassignment_2006,bisgaard_time-resolved_2009} allowing kinetic models to be constructed.
Photoelectron spectra are also sensitive to excited vibrational wavepacket dynamics due to the dependence of the binding energy on the evolving molecular geometry.\cite{hockett_probing_2013,marciniak_interpretation_2014,elkins_isotope_2016} Signatures of such dynamics have been observed in TRPES as shifts\cite{suzuki_time-resolved_2011,fuji_excited-state_2011,horio_ultrafast_2016,hockett_probing_2013,kobayashi_ultrafast_2015,champenois_involvement_2016,horio_full_2016,adachi_probing_2018} in the measured photoelectron kinetic energy as the probe pulse delay is increased. Such shifts are a direct result of the wavepacket motion on the excited state potential energy surface (PES), due to the configuration dependent ionization potential. 

We temporally and spectrally resolve photodissociation dynamics of methanol following excitation to the 2 $^1$A$''$ (S$_2$) electronic state at 156\,nm, using TRPES to identify and follow the  photochemical reaction.
Using a 260\,nm single-photon probe, we measure dynamical features in the photoelectron spectra as a function of the pump-probe time delay, and quantify them to extract signatures of the nuclear motion that emerges in the first few femtoseconds following excitation.
Measurements of the photoelectron kinetic energy $\mbox{E}_{e^-}$ allow the determination of the transient binding energy $\mathcal{E}(\vec{\mathbf{R}})$ of the system, 
\begin{equation}
\begin{split}
	\mathcal{E}(\vec{\mathbf{R}}) &= \hbar\omega_{pump}+\hbar\omega_{probe}-\mbox{E}_{e^-}\\
    &= \hbar\omega_{pump}+\mbox{E}_{c}(\vec{\mathbf{R}})-\mbox{E}_{n}(\vec{\mathbf{R}}),
\label{pbe_eqn}
\end{split}
\end{equation}
where the potential energies of the cation E$_c(\vec{\mathbf{R}})$ and the neutral excited state E$_n(\vec{\mathbf{R}})$ depend on the nuclear coordinates $\vec{\mathbf{R}}$, and E$_c(\vec{\mathbf{R}})$ - E$_n(\vec{\mathbf{R}})$ is the coordinate-dependent vertical ionization energy. The pump and probe photon energies are $\hbar\omega_{pump}$ and $\hbar\omega_{probe}$, respectively.
When interpreted with the aid of electronic structure calculations, which assist in reducing the dimensionality of the dynamics through the identification of a characteristic reaction coordinate, this technique enables the nuclear wavepacket to be tracked as it evolves on the PES. 

Methanol plays an important role in atmospheric chemistry, fuels and energy transport,\cite{olah_beyond_2009,olah_beyond_2005,kothandaraman_conversion_2016} and the excited state dynamics of methanol are highly relevant to each of these applications. Direct investigation of excited state dynamics of methanol in the time domain has been challenging due to the high electronic excitation energies, the ultrafast nuclear motion on the excited states, and the relative complexity of the molecule involving two fundamental functional groups.
In previous studies, several neutral dissociation channels were observed following photoexcitation of the  2$^1$A$''$ (S$_2$) state in methanol.\cite{HarichCompeting,harich_photodissociation_1999,cheng_experimental_2002}
Among these, the dominant two-body breakup processes produce either CH$_3$O+H or CH$_3$+OH, while CH$_2$OH+H and CH$_2$O+H$_2$ are also produced with lower yields.
The S$_2$ excited state is bound along the O--H and C--O coordinates, while the 1$^1$A$''$ (S$_1$) state is dissociative in the same coordinates.\cite{buenker_photolysis_1984}
Previous calculations revealed avoided crossings between the S$_1$ and S$_2$ states along each of these coordinates,\cite{buenker_photolysis_1984} which suggested that the observed dissociation occurs after a nonadiabatic transition to the lower of these two $^1$A$''$ states.\cite{harich_photodissociation_1999,cheng_experimental_2002}
Analysis of the fragment kinetic energy distributions for the O--H and C--O dissociation channels revealed that most of the available energy is channeled into translational kinetic energy, which also supports the possible mechanism of prompt fragmentation on the S$_1$ PES, which connects asymptotically to the CH$_3$O + H and CH$_3$ + OH thermodynamic limits.\cite{harich_photodissociation_1999,cheng_experimental_2002,yuan_photodissociation_2008,philis_resonance-enhanced_2007,chen_product_2011}

In the valence isoelectronic system CH$_3$SH, the S$_1$ and S$_2$ surfaces were previously explored by neutral fragment velocity map imaging\cite{izquierdo_velocity_2006} and {\it ab initio} theory.\cite{stevens_adiabatic_1995,yarkony_role_1994,yarkony_diabolical_1996} In CH$_3$SH, a conical intersection between the S$_2$ and S$_1$ PESs, involving C--S bond stretch and S--H motion, enables dissociation. In the present investigation we explore the possibility of an analogous conical intersection between the S$_2$ and S$_1$ PESs in methanol, and address whether the nuclear dynamics following excitation to the S$_2$ state are mediated by S$_1$, or a previously ignored 1$^1$A$'$ (S$_3$) state.

\section{Methods}
\subsection{Experimental Details}
The experimental setup is described in detail elsewhere\cite{champenois_involvement_2016,shivaram_focal_2016}, therefore only a brief overview is provided here. A 25\,mJ, 25\,fs, 1\,kHz Ti:sapphire near-infrared (NIR) laser system generates vacuum and extreme ultraviolet light via the non-linear process of high-order harmonic generation. The pump and probe pulses are generated by focusing (6~m, f/120) the 780\,nm NIR pulses in a 25~Torr argon gas cell, 10~cm in length, to generate odd harmonics. The fundamental NIR component is then supressed by $>10^{6}$ after the cell, by transmission through a pair of near-grazing incidence dichroic mirrors that have high reflectivity in the vacuum and extreme ultraviolet. The odd harmonics are split into two parallel beams by D-shaped filters and a split mirror interferometer in a back-focusing geometry. In the present experiments, the filters and mirror coatings isolate the fifth harmonic (156\,nm, 7.95\,eV) of the fundamental for the pump arm, and the third harmonic (260\,nm, 4.77\,eV) for delayed probe arm.

The pulse energy partitioning between pump and probe beams at the interferometer was optimized to maximize the relative ionization rate when the pump-probe are coincident, compared to the rate when the probe arrived 50~fs to 150~fs earlier than the pump, and to avoid nonlinear conditions.
After the split mirror interferometer, the foci of the two beams were spatially-overlapped in a velocity map imaging spectrometer, which allowed the kinetic energy spectra of either the electrons or ions, resulting from the photoionization of target molecules, to be recorded.
A set of microchannel plates and a CCD camera captured the projected electron or ion momentum distributions, and the pBASEX algorithm~\cite{garcia_two-dimensional_2004} was used to recover the kinetic energy distributions.
For ions, a fast voltage pulse applied to the detector at specific times relative to the ionizing laser pulse, allowed different ion fragments to be isolated by their mass-to-charge ratio. The global instrument response function was measured to have a gaussian width of 31.2~fs~$\pm$0.9~fs in the time-dependent ion yield for non-resonant two-photon ionization of xenon.
Methanol vapor, from a container held at room temperature, was controlled through a variable leak valve to produce an effusive molecular beam from a long 100\,$\mu$m-diameter capillary, with the capillary exit being 4~cm from the pump and probe foci.

The photoelectron spectra were measured at each pump-probe delay 175 times in a randomized order to minimize artifacts due to small drifts in the laser intensity and target gas pressure. These separate measurements were used to estimate the variance in the signal at each camera pixel for each pump-probe delay. The off-diagonal elements of the resultant covariance matrix were set to zero. These errors were propagated through the background subtraction and pBASEX fit to find the covariance matrix for the time- and energy-dependent photoelectron signal I$(\mathcal{E}$,t). This approach allowed uncertainties in the signal in each energy-time delay bin to be propagated throughout the following analysis.

\subsection{Theoretical Calculations.}
We used equation-of-motion coupled-cluster theory with single and double excitations (EOM-CCSD)~\cite{korona_local_2003,monkhorst_calculation_1977} and an aug-cc-pVTZ basis set~\cite{dunning_gaussian_1989} to calculate the PES of the ground and first few excited states of neutral methanol. The ground state of the cation was calculated using RHF-RCCSD/aug-cc-pVTZ\cite{knowles_coupled_1993,knowles_erratum:_2000}. 
Since methanol's excited states have significant Rydberg character, it is necessary to use augmented functions in the basis set.
At geometries with stretched methanol bonds, we used unrestricted reference functions (UHF-EOM-CCSD)~\cite{valiev_nwchem:_2010,hirata_symbolic_2006,comeau_equation--motion_1993,bartlett_many-body_1978,bartlett_molecular_1980} to ensure that the EOM-CCSD results were reliable and the presence of multireference correlation did not significantly change the results.
The EOM-CCSD calculations were performed using the MOLPRO electronic structure suite\cite{werner_molpro_2012,werner_molpro:_2012}, while those with an unrestricted reference were performed using the NWChem package\cite{valiev_nwchem:_2010,hirata_symbolic_2006}. Conical intersections were discovered using gradient-based searches as implemented in MOLPRO, and by minimizing the energy difference between the two states of interest using a Nelder-Mead simplex method.\cite{nelder_simplex_1965}

Calculations of the two-dimensional potentials in Fig.~\ref{methyl_h} were performed by optimizing the energy on the cation surface with respect to the other internal coordinates. The dimensions C--O--H angle and C--H bond length were chosen partially by comparison to Buenker {\it et al.}\cite{buenker_photolysis_1984}. Additional degrees of freedom were considered, and these two dimensions were chosen to be the most relevant (see the Supporting Information\cite{SI} for details). The geometry optimizations used multireference Rayleigh-Schr\"odinger perturbation theory\cite{werner_third-order_1996,celani_multireference_2000,werner_second_1985,knowles_efficient_1985}, using an active space of 3 electrons in 5 orbitals. Loose geometry convergence parameters were used for some of the geometries far from equilibrium. A single-point energy calculation at the EOM-CCSD and RHF-RCCSD levels for the neutral molecule and cation, respectively, was performed on the optimized geometries.

\section{Results and Discussion}
\begin{figure}
\includegraphics[width=0.45\textwidth]{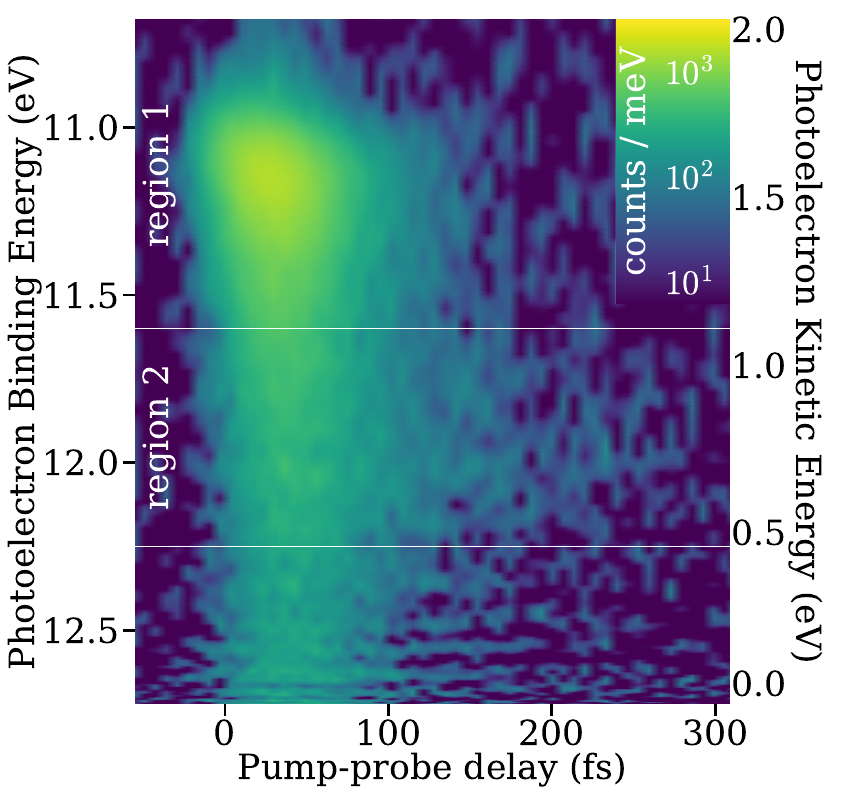}
\caption{\label{TRPES}
Time-resolved photoelectron spectra from ionization at 260\,nm following excitation at 156\,nm, taken in 6.57\,fs steps. The background, due to multiphoton ionization from the pump or probe alone, is removed by subtracting the averaged spectra for the probe pulse earlier than the pump pulse within a time window of t= -50~fs to t=-150~fs. For positive time delays, the probe pulse arrives after the pump pulse. White lines partition the spectra into regions that are discussed in the text.
}
\end{figure}
\begin{figure}
\includegraphics[scale=1]{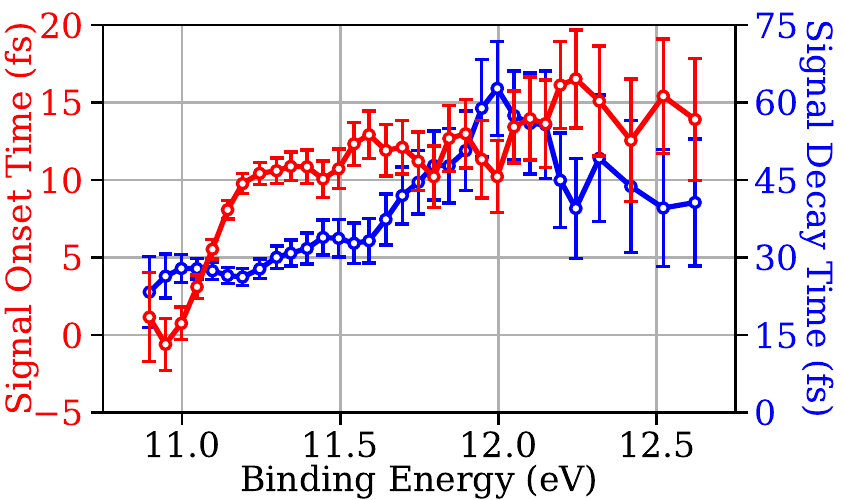}
\caption{\label{decay_fit}
Retrieved photoelectron signal onset times $t_0(\mathcal{E})$ (red) and decay times $\tau(\mathcal{E})$ (blue)
for various binding energy slices extracted from the data of Fig.~\ref{TRPES}, with the error bars representing one standard deviation in the uncertainty of the fit parameters.
The slices are 50\,meV wide, except above 12.25\,eV, where each slice is 100\,meV wide.
}
\end{figure}
\subsection{Signal Onset Times and Resolving a Fast Dissociative Reaction Pathway.}\label{2A}
The experimental time-resolved photoelectron spectra, for excitation at 156\,nm and ionization by a delayed 260\,nm probe pulse, are shown in Fig.~\ref{TRPES}. 
The overall photoelectron yield decays within $\sim$200\,fs. 
We observe features in the photoelectron spectra that shift as the time delay is increased, with a peak first appearing near the binding energy of 10.85\,eV (the ionization potential\cite{macneil_high-resolution_1977}, corresponding to a photoelectron kinetic energy of 1.87\,eV in the present experiments) and extending towards higher binding energies (lower kinetic energies) with longer time delays. 
The probe pulse intensity and photon energy are carefully selected to limit access to excited electronic states of the cation, so that the TRPES features are due to ultrafast wavepacket motion on the neutral excited states and the topology of the ground state PES of the cation.

To quantify the measured photoelectron spectra, we use a global fitting algorithm that fits each energy-slice, $I(\mathcal{E},t)$, with a function parameterized by the energy-dependent signal onset time $t_0(\mathcal{E})$ and exponential decay time $\tau(\mathcal{E})$, convolved with a global Gaussian function $g$ of width $\sigma$:
\begin{multline}
    I(\mathcal{E},t)=A(\mathcal{E})\left
    [H(t-t_0(\mathcal{E}))\times\text{exp}\left(-\frac{t-t_0(\mathcal{E})}{\tau(\mathcal{E})}\right)\right ]\\
    \otimes g(t-t_0(\mathcal{E});\sigma),
\label{fit_eqn}
\end{multline}
where H(t) is the Heaviside step function\cite{champenois_involvement_2016}.
The retrieved signal onset times, $t_0(\mathcal{E})$, and signal decay times, $\tau(\mathcal{E})$, are shown in Fig.~\ref{decay_fit}. Eq. \ref{fit_eqn} provides an energy-dependent framework to the TRPES technique to allow access to time-resolved dynamics within the duration of pump and probe pulses. Further details, including a comparison between this approach and decay associated spectra\cite{townsend_b21u+1_2006} can be found in the Supporting Information\cite{SI}. 

The retrieved onset times can be interpreted as the time it takes for the wavepacket to arrive at a region of the PES with a specified binding energy.
Following photoexcitation, the excited state wavepacket rapidly moves away from Franck-Condon geometries to a region of increased binding energy. 
Over the first $\sim$10\,fs, the  binding energy increases smoothly at 0.05\,eV/fs.
Above 11.25\,eV, there is an abrupt change in this rate to 0.3\,eV/fs, which continues for the remainder of the measurable binding energies up to 12.72\,eV, suggesting neutral dissociation.
Thus, by analyzing the time-dependence of the excited state binding energies (see onset times of Fig.~\ref{decay_fit}), the initial motion of the excited molecule can be accessed on timescales significantly shorter than the $\sim$20~fs widths of the pump and probe pulses.

In methanol, there are 12 internal degrees of freedom.
While this number is small relative to other alcohols, examining excited state dynamics in methanol is challenging due to the high dimensionality of its PESs.
In the present work, we use equation-of-motion coupled-cluster theory  to examine cuts of the PESs of the involved excited states. We define a 1-dimensional coordinate, using the following procedure, to significantly reduce the dimensionality of the system.

A comparison can be made between experiment and theory by considering the photoelectron binding energies. The experiment measures the signal onset times as a function of the binding energy t$_0(\mathcal{E})$, while the theory predicts the binding energies for an electronic state as a function of nuclear geometry $\mathcal{E}(\vec{\mathbf{R}})$. If the dimensionality of nuclear geometries can be reduced through, for example, the definition of a one-dimensional reaction coordinate, the combination of the measurement and calculations allows for the evolving molecular geometry along this reaction coordinate to be tracked. Inversion of the dependent variables for both of the above quantities gives $\mathcal{E}(t)$ and $\vec{\mathbf{R}}(\mathcal{E})$, which can be combined to $\vec{\mathbf{R}}(t)=\vec{\mathbf{R}}(\mathcal{E}(t))$. Both the C--O stretch and C--O--H bond angle were considered, following the measured rates for methanol and the deuterated species presented in Section~\ref{2B}, and the earlier study of Buenker {\it et al.}\cite{buenker_photolysis_1984}. 

\begin{figure}
\includegraphics[width=0.95\linewidth]{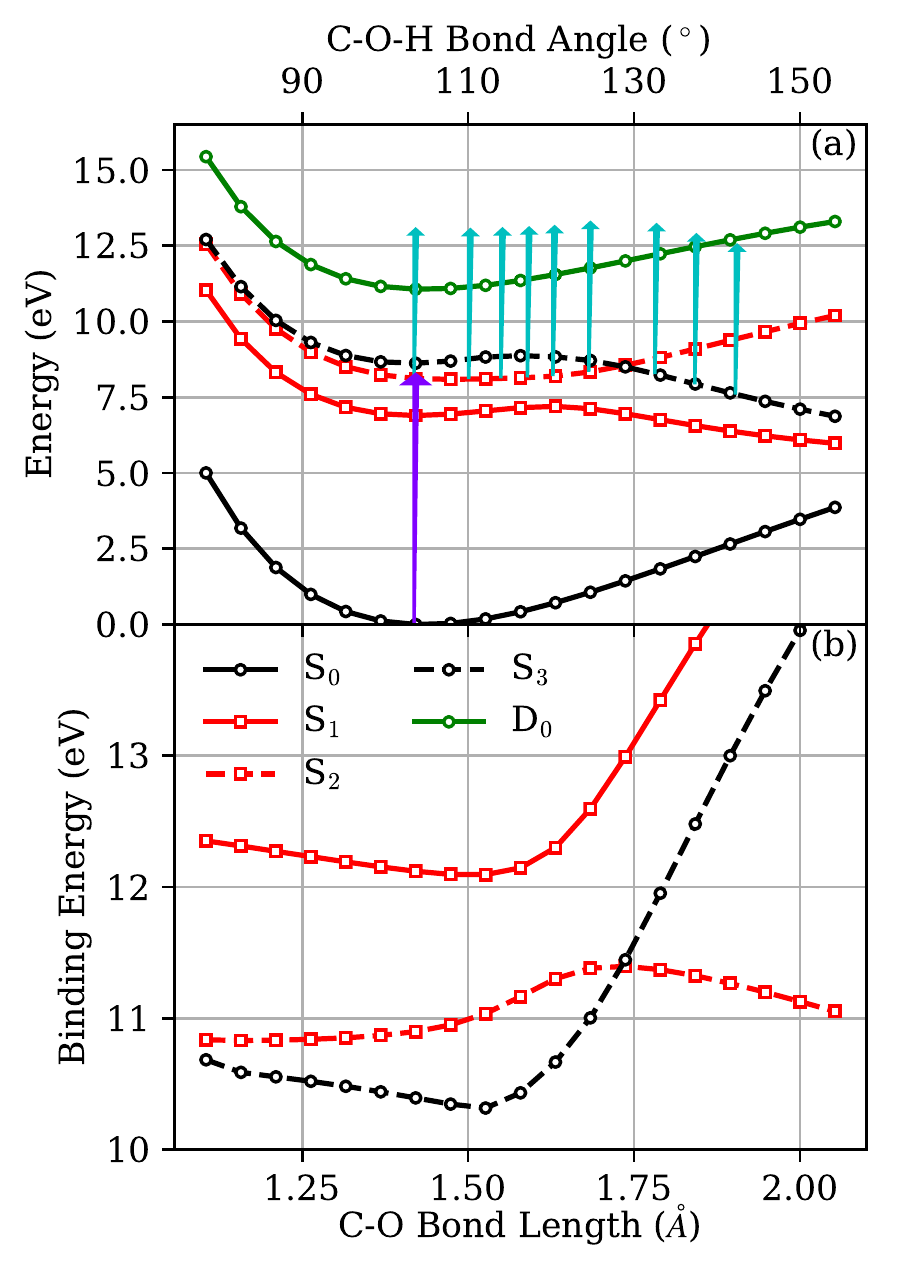}
\caption{\label{PES_CO}
Calculated cuts (a) of the potential energy surfaces and (b) the associated photoelectron binding energies for the first two A$'$ states (black lines, S$_0$ and S$_3$), first two A$''$ states (red lines, S$_1$ and S$_2$), and cation ground state (green line, D$_0$) as the C--O bond distance and C--O--H bond angle are simultaneously stretched and opened, with arrows indicating the initial excitation (magenta) and time-delayed ionization (cyan).
The ionization is shown in 2\,fs steps starting at pump-probe overlap, using the experimental  binding energy dependent onset times of Fig.~\ref{decay_fit} and the  binding energy calculations shown here, and making the approximation that all of the measured  binding energy shifts are due to motion along this coordinate.}
\end{figure}

To explain the measured photoelectron spectral shifts, we performed this analysis on the simultaneous stretching of the C--O bond and the opening of the C--O--H bond angle, to identify a reaction coordinate representative of the global topology of S$_2$, as shown in Fig.~\ref{PES_CO}.
The initial excited wavepacket will advance on the S$_2$ (2$^1$A$''$) PES along this coordinate until it reaches a region of nonadiabatic coupling between this surface and that of S$_3$ (1$^1$A$'$).
At and in the neighborhood of this geometry, internal conversion to S$_3$ becomes possible.
Starting at this geometry on S$_3$, the same fragments, specifically those due to C--O and O--H cleavage, can be reached as previously considered~\cite{harich_photodissociation_1999,cheng_experimental_2002}. The S$_3$ and S$_1$ (1$^1$A$''$) surfaces become degenerate at linear C--O--H geometries, however crossing to S$_1$ is not required for dissociation, in contrast to the valence isoelectronic system CH$_3$SH, and previous assumptions about methanol.

The calculations along this reaction coordinate agree well with the features of the signal onset time measurement.
At 11.25~eV the abrupt change in the gradient of the measured binding energy-dependent onset time (Fig.~\ref{decay_fit}) is consistent with a transition from the initially excited S$_2$ to S$_3$, where the  binding energy depends more strongly on the nuclear geometry.
The energetic location of this transition is also in good agreement with the calculated crossing geometry of these two states at $\mathcal{E}(\vec{\mathbf{R}})=$ 11.4\,eV, as seen in Fig.~\ref{PES_CO}.
After the internal conversion to the dissociative state, the other coordinates begin to play a role.
Notably, the O--H bond breaking on S$_3$ is expected to have a steep gradient compared to the other dissociation channels~\cite{marston_product_1993}.

The measured signal decay times of Fig.~\ref{decay_fit} characterize the timescale for the excited state wavepacket to exit a region of the PES at a specified binding energy. For binding energies associated with S$_2$, the decay times are between 25 and 30\,fs, which is consistent with the $\sim$26\,fs lifetime that can be extracted from the $\sim$100\,meV wide vibrational structure seen in the S$_2$ absorption spectrum at 156\,nm~\cite{sominska_absorption_1996,cheng_experimental_2002,varela_excitation_2015}. In this case, the measurements reveal the time needed to completely depopulate the Franck-Condon region, augmenting the information provided by the signal onset times. 

The previous studies of this excited state of methanol consider only the crossing between the two A$''$ states S$_1$ and S$_2$~\cite{buenker_photolysis_1984,harich_photodissociation_1999}.
A calculation on CH$_3$SH~\cite{yarkony_role_1994} suggests that an analogous conical intersection could be arrived at largely through C--S (analogous to C--O) bond stretch.
We performed a search for a conical intersection between S$_1$ and S$_2$ in methanol, and only found one in a region of the PES that is energetically inaccessible for the present experiments. The importance of the A$'$ state S$_3$ in the dissociation of methanol underlines the weak correspondence in the excited state dynamics of the two isoelectronic molecules. The previously ignored S$_3$ state appears to be responsible for the rapid relaxation within 15\,fs observed in the present experiments, and is also consistent with previous measurements.\cite{buenker_photolysis_1984,harich_photodissociation_1999}

\subsection{Correlating Delayed-Onset Photoelectrons with Hydrogen Elimination.}\label{2B}
\begin{figure}
\includegraphics[scale=1]{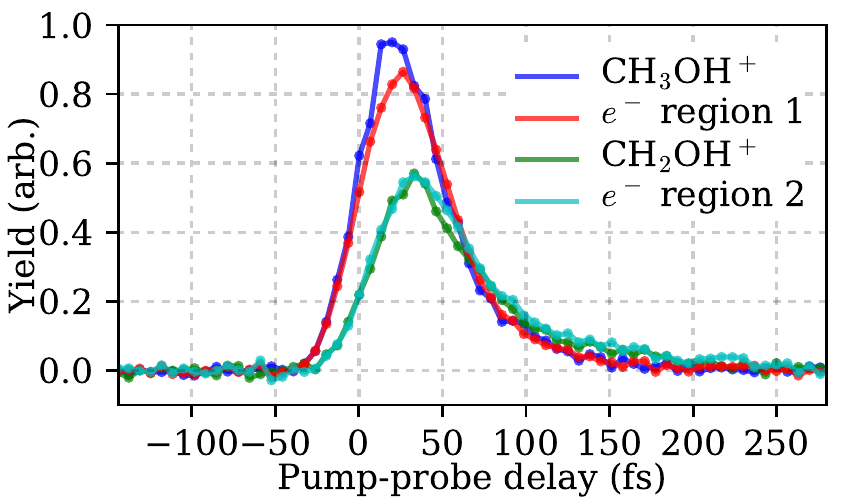}
\caption{\label{ch3oh}
Pump-probe time-dependent yield of the two ion and photoelectron channels.
In this analysis, only the total ion and photoelectron yields were normalized.
The two electron regions correspond to 10.85 to 11.6\,eV (red, fast decay) and 11.6 to 12.25\,eV (cyan, slow decay), as labeled in Fig.\,\ref{TRPES}.}
\end{figure}

Excited methanol fragments through multiple dissociation channels following photoexcitation.
Both the CH$_2$OH$^+$+H channel and the CH$_2$O$^+$+H$_2$ channel, with appearance energies of 11.6\,eV and 12.45\,eV, respectively, are energetically accessible for the photon energies of the present experiments.
We measured the time-dependent ion yields for the species associated with these H and H$_2$ elimination channels along with the bound ionic methanol channel using the same experimental apparatus and the same 156\,nm pump, 260\,nm probe scheme.
The results are shown in Fig.~\ref{ch3oh}.
A fast rise in the CH$_3$OH$^+$ yield near $t_0$ with a fast decay, and a slightly delayed rise in the CH$_2$OH$^+$ yield with a slower decay, are observed.
Hydrogen loss in the latter channel was confirmed to occur exclusively from the methyl site of methanol by performing the same experiment with its deuterated isotopologues CD$_3$OH and CH$_3$OD.
There was no significant time-dependent yield of CH$_2$O$^+$ measured in the present experiments.

Fitting the time-dependent ion yields with the ansatz of Eq.~\ref{fit_eqn} gives a 7\,fs delay for the fragment signal relative to that of the parent ion and a 30\,fs and 45\,fs $1/e$ lifetime for the parent and fragmented ion, respectively.
Although these ion yield measurements were not made in coincidence with the photoelectron measurements, the faster CH$_3$OH$^+$ yield correlates well with that of photoelectrons with  binding energies between the ionization potential and 11.6\,eV (region 1 in Fig.~\ref{TRPES}), as shown in Fig.~\ref{ch3oh}, and the slower CH$_2$OH$^+$ yield correlates with that of photoelectrons between 11.6\,eV and 12.25\,eV (region 2 in Fig.~\ref{TRPES}), where the signal takes longer to decay (Fig.~\ref{decay_fit}). 
Note that the time-dependence in the fragment ion yield and the correlated photoelectron yield exhibits later onset times followed by longer decay times. 
The fitted signal decay lifetimes for the fragment and parent ion yields, along with the corresponding photoelectron yields from each of the binding energy bands described above, are summarized in Table~\ref{lifetimes} for each target species.
Although C--H fragmentation occurs on the methyl site of the molecule, the signal decay timescales were greatly affected by hydroxyl deuteration and less so by fully deuterating the methyl functional group, indicating that hydroxyl motion in the excited state plays an important role in facilitating the methyl hydrogen elimination. A previous experimental study of 157~nm photodissociation\cite{harich_photodissociation_1999} determined two-body methyl hydrogen elimination to be a minor photodissociation channel (3\% compared to two-body O--H break), therefore significant C--H break is not expected to occur on the neutral excited states, but may occur on the ground state of the cation.
Deuteration did not strongly affect the appearance time of the fragment ion, relative to the parent ion, nor the photoelectron signal onset times, indicating that stretching of the C--O bond, rather than hydrogen motion, limits the rate of the initial dynamics.

\begin{table}
\caption{\label{lifetimes}
Fitted exponential decay times, in femtoseconds, of the ion and electron channels presented in Fig.\,\ref{ch3oh} and in the text, for methanol and its deuterated isotopologues, with uncertainties representing one standard deviation.
The two electron regions are illustrated in Fig.\,\ref{TRPES}.
NM indicates a channel that was not measured.}
\begin{ruledtabular}
\begin{tabular}{cccc}
Channel&CH$_3$OH&CH$_3$OD&CD$_3$OH\\
\hline
Parent ion&30$\pm$3&43$\pm$6&36$\pm$6\\
$e^-$ region 1&29$\pm$2&39$\pm$9&NM\\
Fragment ion&45$\pm4$&91$\pm$8&56$\pm$6\\
$e^-$ region 2&48$\pm$4&97$\pm$16&NM
\end{tabular}
\end{ruledtabular}
\end{table}

\begin{figure}
\includegraphics[width=0.95\linewidth]{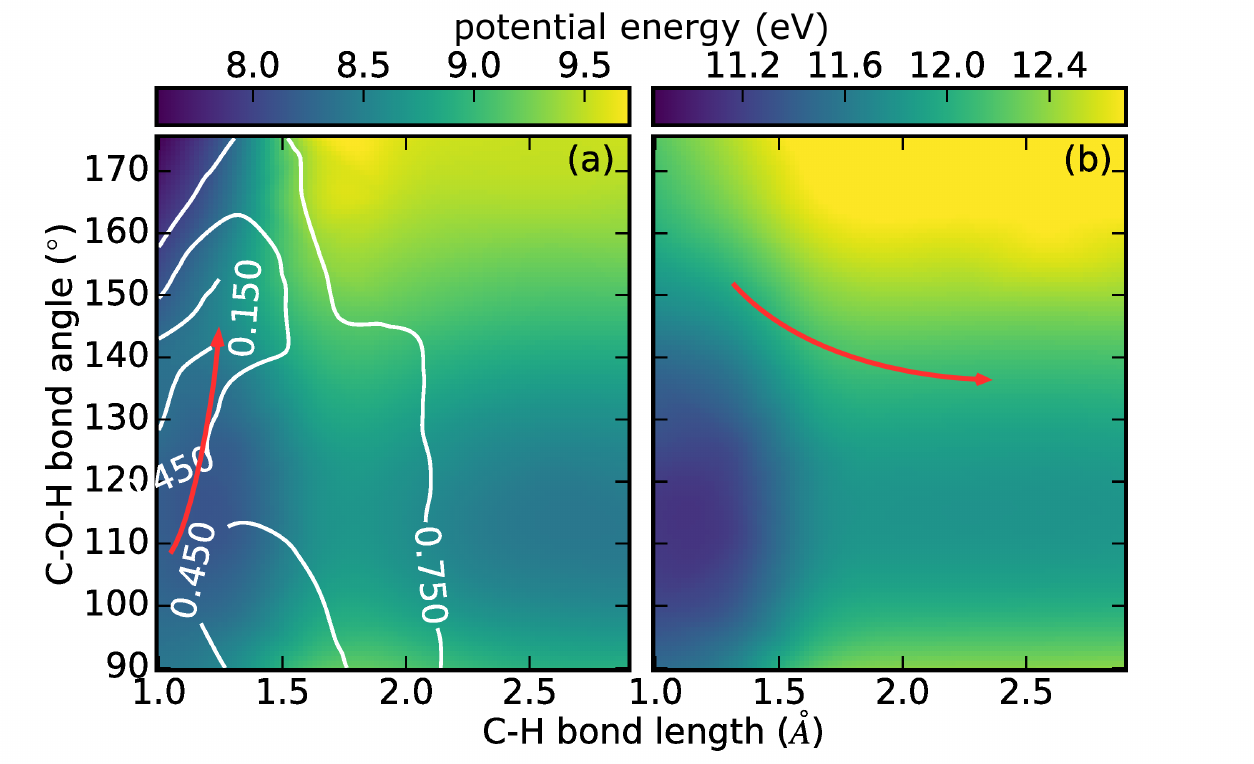}
\caption{\label{methyl_h}
2-dimensional cuts of (a) the S$_2$ and S$_3$ excited state PES, where the lower of the two surfaces is illustrated by the color map (energy scale in eV) and the difference $|$S$_3$-S$_{2}|$ by white contours, and (b) the cation ground state D$_0$ PES (energy scale in eV). 
The excited state wavepacket in the Franck-Condon region of S$_3$ experiences a potential favoring C--O--H angle opening, leading to a conical intersection between S$_3$ and S$_2$, as indicated by the red arrow on the left panel (a). A C--H dissociation path on the cation surface D$_0$, indicated by a red arrow on the right panel (b), may follow ionization of the excited neutral system by a single probe photon, for open C-O-H angles $\agt$ 140$^\circ$ and pump-probe time delays between 10~fs and 200~fs.
}
\end{figure}

To gain insight into the dynamics leading to methyl hydrogen loss in the present experiments, we turn to the computed 2-dimensional cuts, along the C--H stretch and C--O--H angle coordinates, of the S$_2$ and S$_3$ excited states and the cation ground state D$_0$, which are shown in Fig.\,\ref{methyl_h}. The cuts were chosen following the results of Buenker {\it et al.}~\cite{buenker_photolysis_1984} and the observed increase in decay time with hydroxyl deuteration, both of which showed this bond angle to play a key role. Other degrees of freedom were considered, for instance a linear interpolation between reactants and products with orthogonal degrees optimized, and we determined the most relevant information to be in Fig.~\ref{methyl_h} (see the Supporting Information\cite{SI} for a discussion of the other degrees of freedom). 
The cuts reveal a path for a fraction of the cation wavepacket on D$_0$ to dissociate by C--H bond cleavage. The pathway is reached following ionization by the probe photon for C--O--H angles $\agt 140^\circ$, which are accessed through a conical intersection between the S$_2$ and S$_3$ states. A low potential energy barrier ($<$ 400\,meV) on S$_2$ to angle opening is consistent with the delayed onset of the CH$_2$OH$^+$ fragment ion and the correlated lower-energy photoelectrons in the present measurements on CH$_3$OH. Such a barrier also provides a possible explanation for the longer CH$_2$OD$^+$ fragment ion and photoelectron decay times observed in the present measurements on deuterated methanol CH$_3$OD. Significant barriers around 1.7\,\AA\, on both of the S$_2$ and S$_3$ cuts may inhibit direct C--H dissociation by 156\,nm photoexcitation, as observed in previous experimental investigations\cite{harich_photodissociation_1999}, however motion along other coordinates reduce the barriers.

\section{Conclusions}

We have measured and described the ultrafast dynamics of methanol after 156\,nm photoexcitation, and demonstrated sensitivity to the coupled electronic and nuclear motion immediately following electronic excitation, using the analysis of onset and decay times in time-resolved photoelectron spectral features.
By calculating cuts through the potential energy surfaces of the electronic excited states, we find mechanisms to explain the measured dissociation processes, invoking a previously unexplored A$'$ electronic state S$_3$.
This state is accessed nonadiabatically through a conical intersection, and dissociates rapidly along the C--O bond.
The nonadiabatic transition from the photoexcited S$_2$ state to the dissociative S$_3$ state can occur within 15\,fs of excitation.
C--H dissociation was observed to be delayed relative to the C--O fragmentation process and was determined to occur on the ground electronic state of the methanol cation, following significant C--O--H angle opening, facilitated by the conical intersection between the neutral states S$_3$ and S$_2$.  

Comparison of the time-resolved measurements with the computed PES cuts enables the identification of the relevant reaction coordinates and a reduction in dimensionality of the dynamics.
Resolving nuclear motion in higher dimensions through this procedure could be possible in further experiments involving photoionization to multiple electronic continua, given that the valence electron binding energies have unique dependencies on molecular configuration. 
These approaches could enable a more detailed understanding of nonadiabatic dynamics in electronically excited polyatomic molecules, and may also be useful in studying highly coupled electronic nuclear motion, even on sub-10\,fs timescales~\cite{calegari_ultrafast_2014,vacher_electron_2017}.

\begin{acknowledgments}
This research used resources of the National Energy Research Scientific Computing Center, a United States Department of Energy (US DOE) Office of Science User Facility supported by the Office of Science of the DOE, and was supported by the DOE, Office of Science, Office of Basic Energy Sciences, Chemical Sciences, Geosciences, and Biosciences Division, under Contract No. DE-AC02-05CH11231. Work at the University of California, Davis was supported by the US Army Research Laboratory and the US Army Research Office under Grant No. W911NF-14-1-0383. 
\end{acknowledgments}

\bibliography{methanol}

\end{document}